\documentclass[10pt,twocolumn]{article}
\usepackage{arxiv}
\usepackage{times}
\usepackage{epsfig}
\usepackage{graphicx}
\usepackage{amsmath}
\usepackage{amssymb}
\usepackage[pagebackref=true,breaklinks=true,letterpaper=true,colorlinks,bookmarks=false]{hyperref}
\usepackage{booktabs}
\usepackage{multirow, makecell}

\usepackage{flushend}
\usepackage{ctable}

\usepackage{soul}
\usepackage{amsmath}
\usepackage{amssymb}
\usepackage{microtype}
\usepackage{wrapfig}

\newcommand{\refFig}[1]{Fig.~\ref{fig:#1}}

\soulregister\ref7
\soulregister\cite7
\soulregister\refFig7
\soulregister\cite7
\soulregister\ref7
\soulregister\pageref7
\soulregister\shortcite7
\soulregister\eg7
\soulregister\ie7
\soulregister\etal7
\soulregister\reftbl7

\DeclareGraphicsExtensions{.png,.jpg,.pdf,.ai,.psd}
\newcommand{\mysection}[2]{\section{#1}\label{sec:#2}}
\newcommand{\mysubsection}[2]{\subsection{#1}\label{sec:#2}}

\usepackage{pifont}
\usepackage{adjustbox}

\newcolumntype{R}[2]{%
    >{\adjustbox{angle=#1,lap=\width-(#2)}\bgroup}%
    l%
    <{\egroup}%
}\newcommand
*\rot[2]{\multicolumn{1}{R{#1}{#2}}}%

\begin{document}

\title{LRG at TREC 2020: Document Ranking with XLNet-Based Models}


\author{
  \textbf{Abheesht Sharma}\\
 \textit{BITS Pilani, KK Birla Goa Campus}\\
  \texttt{\small f20171014@goa.bits-pilani.ac.in} \\
   \and
 \textbf{Harshit Pandey} \\
  \textit{AISSMS IOIT, SPPU}\\
  \texttt{\small hp2pandey1@gmail.com} \\
}

\pagestyle{plain}
\maketitle

\begin{abstract}
Establishing a good information retrieval system in popular mediums of entertainment is a quickly growing area of investigation for companies and researchers alike. We delve into the domain of information retrieval for podcasts. In Spotify’s Podcast Challenge, we are given a user’s query with a description to find the most relevant short segment from the given dataset having all the podcasts. Previous techniques that include solely classical Information Retrieval (IR) techniques, perform poorly when descriptive queries are presented. On the other hand, models which exclusively rely on large neural networks tend to perform better. The downside to this technique is that a considerable amount of time and computing power are required to infer the result. We experiment with two hybrid models which first filter out the best podcasts based on user's query with a classical IR technique, and then perform re-ranking on the shortlisted documents based on the detailed description using a transformer-based model.
\end{abstract}

\mysection{Introduction}{Introduction}
Podcasts are exploding in popularity. With the addition of the DIY podcasting platform Anchor, everyone today has access to tools to create their own podcasts and publish it to Spotify, and hence, the landscape is growing ever richer and more diverse. As the medium grows, a very interesting problem arises: how can one connect one's users to podcasts which align with one's interest? How can a user find the needle in the haystack and how can the user be presented with a list of potential podcasts customized to their interests? The Spotify podcasts track of TREC 2020 attempts to enhance the discoverability of podcasts with a task that attempts to retrieve the jump-in point for relevant segments of podcast episodes given a query and a description.

The dataset consists of 100,000 episodes from a large variety of different shows on Spotify. The sampled episodes range from amateur to professional podcasts including a wide variety of formats, topics and audio quality. Each episode consists of a raw audio file, an RSS header containing the metadata (such as title, description and publisher) and an automatically-generated transcript in English that boasts a word error rate of just 18.1\% as well as a named entity recognition rate of 81.8\%.

Traditional information retrieval techniques use statistical methods such as TF-IDF to score and retrieve relevant segments. While these methods work well on straightforward queries which consist of only a few terms, and which can be easily matched with the keywords of the document to be retrieved, the more complicated queries with abstract questions will end up failing in traditional IR techniques. The Spotify dataset \cite{spotify} for the information retrieval task consists of two components for a user's search: the query and a description along with the query, which is a more verbose query highlighting the specific requirements of the user. A traditional IR system ends up performing poorly if we consider the descriptive query. On the other hand, transformer-based models like BERT have been used for information retrieval tasks and have achieved satisfactory results considering both the parameters of the user's search. However, the amount of time required to process one search request is considerably more than the traditional IR methods.

Our approaches use both: a traditional information retrieval system as well as a transformers-based model. For the traditional information retrieval model, we use a combination of BM25 \cite{bm25}, a traditional relevance ranking model, as well as, RM3 \cite{rm3}, a relevance-based language model for query expansion. We filter the top thousand podcasts from this IR model and pass it to our transformer-based model. For the transformer-based model, we use XLNet \cite{XLNet}, a Permutation Language Model (PLM) with two different modifications. For the first type of XLNet model, we add a simple linear layer to XLNet that performs a regression task to generate a score reflecting how relevant the query-document pair is.

The second approach is contextualised re-ranking with XLNet. We use XLNet to compute the query and document embeddings separately, and we use these embeddings to compute the similarity matrix between the query and the document, followed by kernel pooling techniques to arrive at a relevance score. The advantage of this method is that we can create, store and index the embeddings of documents for future use. Storing contextualised representation makes the query-inference time much lesser than the above mentioned regression model, according to Sebastian Hofstätter et al.\cite{SimpleContextualisation}.

\mysection{Related Work}{Related Work}
In the earlier days, before the advent of neural networks, people relied on statistical or probabilistic algorithms such as TF-IDF, BM25 \cite{bm25} and RM3 \cite{rm3}. BM25 is based loosely on the TF-IDF \cite{tf_idf} model but improves on it by introducing a document length normalization term and by satisfying the concavity constraint of the term frequency (taking logarithm of the term frequency, etc.). RM3 is an association-based and relevance-based language model, useful for query expansion. BM25 and RM3 together form a lethal combination (namely, BM25+RM3), and is reliable and efficient.

However, with the meteoric rise of neural networks (and the fact that BM25+RM3 is essentially a statistical method), researchers started looking for models which can learn unique representations for words based on their context, usage in text, etc. A few CNN/RNN-based methods are DRMM \cite{DRMM}, DSSM \cite{DSSM}, CDSSM \cite{CDSSM} and K-NRM \cite{KNRM}. 

One of the earliest attempts in neural research for document retrieval was with DSSM \cite{DSSM}. The DSSM model ranked the query and document simply by their representation similarities. Later on, a version of DSSM with CNNs was introduced, also known as CDSSM.\cite{CDSSM}.

DRMM, a CNN-based model, introduced a histogram pooling technique to summarize the translation matrix and hence showed that counting the word level translation scores at different soft match levels is more effective than weight-summing them.

K-NRM \cite{KNRM} uses the similarities between the words of the query and the document to build a translation matrix and then uses kernel-pooling to obtain summarised word embeddings and at the same time, provides soft-match signals in order to learn how to rank. The Conv-KNRM \cite{CONV_KNRM} model uses a CNN for soft-matching of n-grams. CNNs are used to represent n-grams of various lengths which are then soft-matched in a unified embedding space. It ranks using n-grams soft matches with Kernel Pooling and learning to rank layer.

Then came the era of transformer-based information retrieval approaches. One of the first transformer based methods was using BERT for ad hoc-retrieval \cite{BERTforIR}. BERT is used to get a combined representation of the query and the document with a linear layer on top which is then used for obtaining a score. XLNet \cite{XLNet} follows a permutation language modelling approach that outperforms BERT on various NLP tasks, which is why we we chose XLNet as the encoder.

Hofstätter, et. al. \cite{SimpleContextualisation} combine the transformers with the KNRM method. The difference from the KNRM approach is that they use an n-layered transformer to obtain embeddings for the query and the document.

\mysection{Our Approach}{OurApproach}

\begin{figure}
    \includegraphics[width=0.6\textwidth]{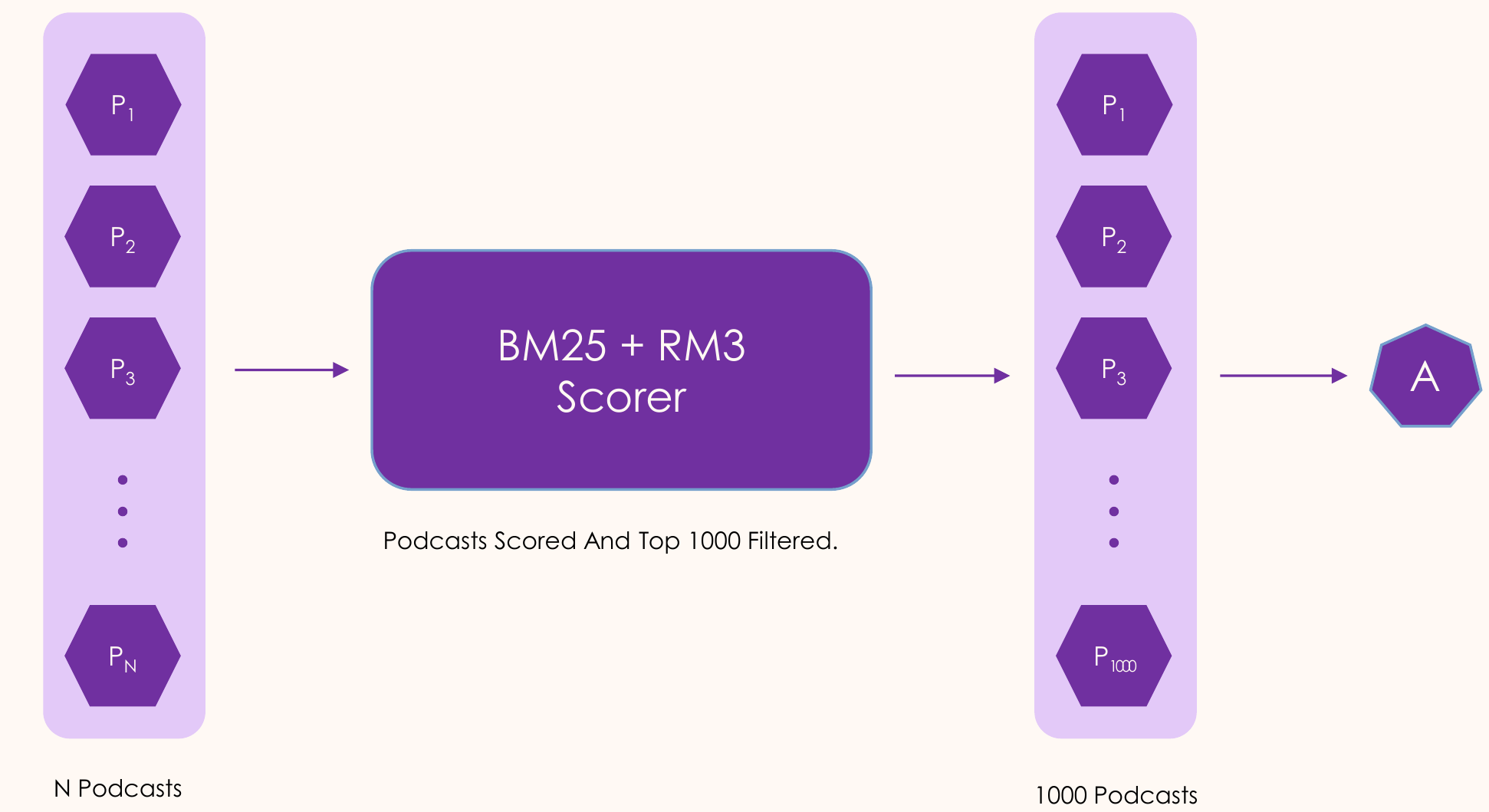}
    \caption{Shortlisting podcasts using BM25 and RM3}
\end{figure}

\begin{figure*}
    \includegraphics[width=1\textwidth]{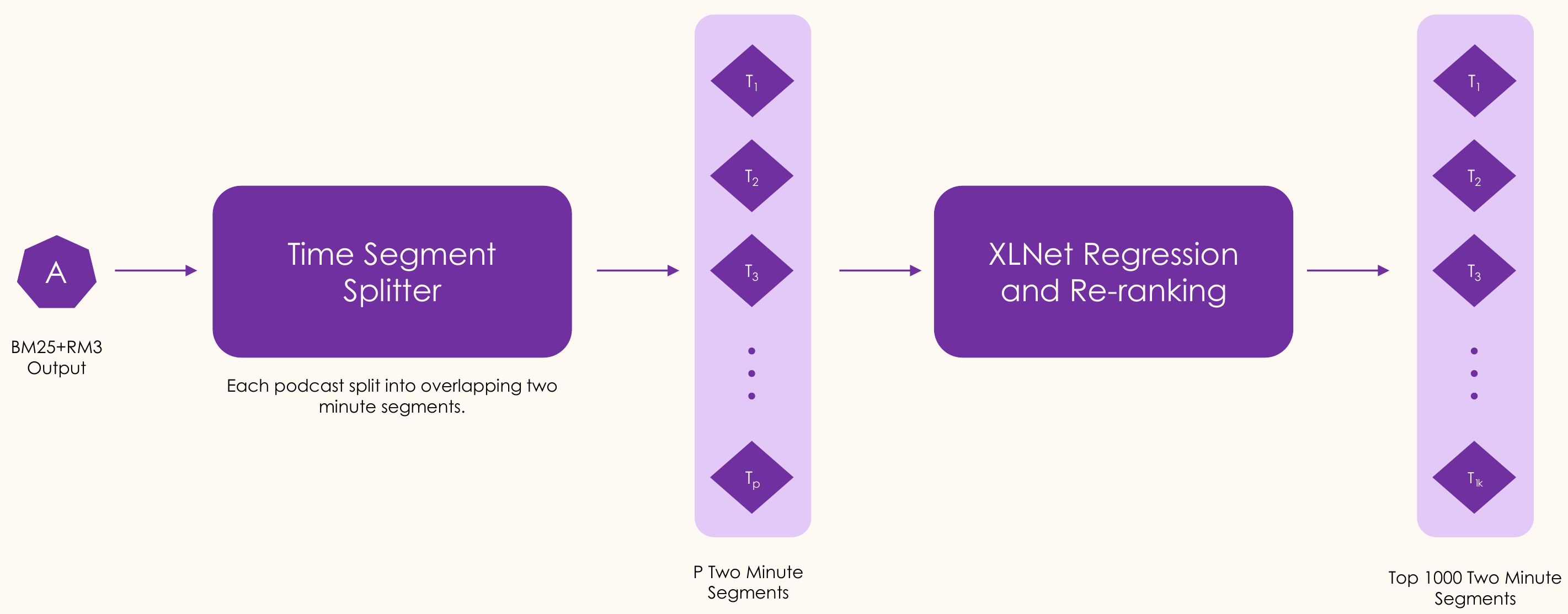}
    \caption{Splitting podcasts into two minute segments and reranking them with XLNet.}
\end{figure*}
Here, We elucidate the algorithms and models that are required to implement our approach. We start off with the traditional IR method, i.e., BM25 and RM3, and then, move on to XLNet Regression and XLNet with Similarity. 

\mysubsection{BM25}{BM25}

BM25 \cite{bm25} is a bag-of-words retrieval function which ranks documents based on the appearance of query terms in every document, not taking into consideration the proximity of the query words within the document. The general instantiation method is as follows:

Given a query Q, containing keywords ${q_{1},...,q_{n}}$ the BM25 score of a document D is:
\begin{equation}
\operatorname{score}(D, Q)=\sum_{i=1}^{n} \operatorname{IDF}\left(q_{i}\right) \cdot \frac{f\left(q_{i}, D\right) \cdot\left(k_{1}+1\right)}{f\left(q_{i}, D\right)+k_{1} \cdot\left(1-b+b \cdot \frac{|D|}{\operatorname{avgdl}}\right)}\\
\end{equation}
where $f\left(q_{i}, D\right)$ is $q_{i}$ 's term frequency in the document $D,|D|$ is the length of the document $D$ in words, and $avgdl$ is the average document length amongst other documents from the text collection. $k_{1}$ and $b$ are free parameters, usually chosen, in absence of an advanced optimization, as $k_{1} \in[1.2,2.0]$ and $b=0.75$.  $\mathrm{IDF}\left(q_{i}\right)$ is the IDF (inverse document frequency) weight of the query term, $q_{i}$. It is usually computed as:
\begin{equation}
\operatorname{IDF}\left(q_{i}\right)=\ln \left(\frac{N-n\left(q_{i}\right)+0.5}{n\left(q_{i}\right)+0.5}+1\right)
\end{equation}
 where $N$ is the total number of documents in the collection, and $n(q_i)$ is the number of documents containing $q_i$.
\mysubsection{RM3}{RM3}

In the first estimation method of relevance model (often called RM1), the query likelihood $p(Q \mid D)$ is used as the weight for document $D$. For every word $w$, we average over the probabilities given by each document language model. The formula of RM1 is:
\begin{equation}
p_{1}(w \mid Q) \propto \sum_{\theta_{D} \in \Theta} p\left(w \mid \theta_{D}\right) p\left(\theta_{D}\right) \prod_{i=1}^{m} p\left(q_{i} \mid \theta_{D}\right)
\end{equation}
where $\Theta$ denotes the set of smoothed document models in the pseudo feedback collection $F$, and $Q=\left\{q_{1}, q_{2}, \cdots, q_{m}\right\}$.

In RM2, the term $p(w | Q)$ is computed using document containing both query terms and word.
\begin{equation}
p_{2}(w \mid Q) \propto p(w) \prod_{i=1}^{m} \sum_{\theta_{D} \in \Theta} p\left(q_{i} \mid \theta_{D}\right) \frac{p\left(w \mid \theta_{D}\right) p\left(\theta_{D}\right)}{p(w)}
\end{equation}

RM3 is based on RM1 and RM2, and it uses the Dirichlet Smoothing Method to smooth the language model of each pseudo-relevant document. To enhance the performance, a linear combination of $P(w \mid Q)$ and $\theta_{Q}$ can be taken \cite{PseudoRelevance}.
\begin{equation}
\mathrm{RM} 3: p\left(w \mid \theta_{Q}^{\prime}\right)=(1-\alpha) p\left(w \mid \theta_{Q}\right)+\alpha p_{1}(w \mid Q)
\end{equation}
We chose to use RM3 \cite{rm3} over RM4 and other query language models because of the conclusions derived from \cite{PseudoRelevance}.

\mysubsection{BM25 + RM3}{BM25 + RM3}
We score each episode using BM25+RM3 model and hence select the top 1000 episodes to be processed further. Compared to QL, QL+RM3 and other combinations, performed the best \cite{bm25_rm3_comparison}.

\mysubsection{XLNet Regression}{XLNet Regression}

The XLNet paper consolidates the latest advances in NLP research with inventive decisions in how the language modelling problem is approached. XLNet achieves state-of-the-art for a multitude of NLP tasks. XLNet is a Permutation Language Model. It calculates the probability of a word token given all permutations of word tokens in a sentence, instead of just those before or just after the target token, i.e., it takes bidirectional context into account.

Previous works have used BERT. However, since the online implementations of BERT have a limit on the sequence length (up to 512 tokens), and XLNet can handle large documents (it has no token limit), we proceed with XLNet.
The formal definition of the XLNet modeling objective is as follows:
\begin{equation}
\max _{\theta} \mathbb{E}_{\mathbf{z} \sim \mathcal{Z}_{T}}\left[\sum_{t=1}^{T} \log p_{\theta}\left(x_{z_{t}} \mid \mathbf{x}_{\mathbf{z}<t}\right)\right]
\end{equation}
where $\mathcal{Z}_{T}$ is the set of all possible permutations of the length- $T$ index sequence $[1,2, \ldots, T]$. The use of $z_{t}$ and $\mathbf{z}_{<t}$ is to denote the $t$ -th element and the first $t-1$ elements of a permutation $\mathrm{z} \in \mathcal{Z}_{T} .$ \cite{XLNet}.
\begin{figure*}
    \includegraphics[width=1\textwidth]{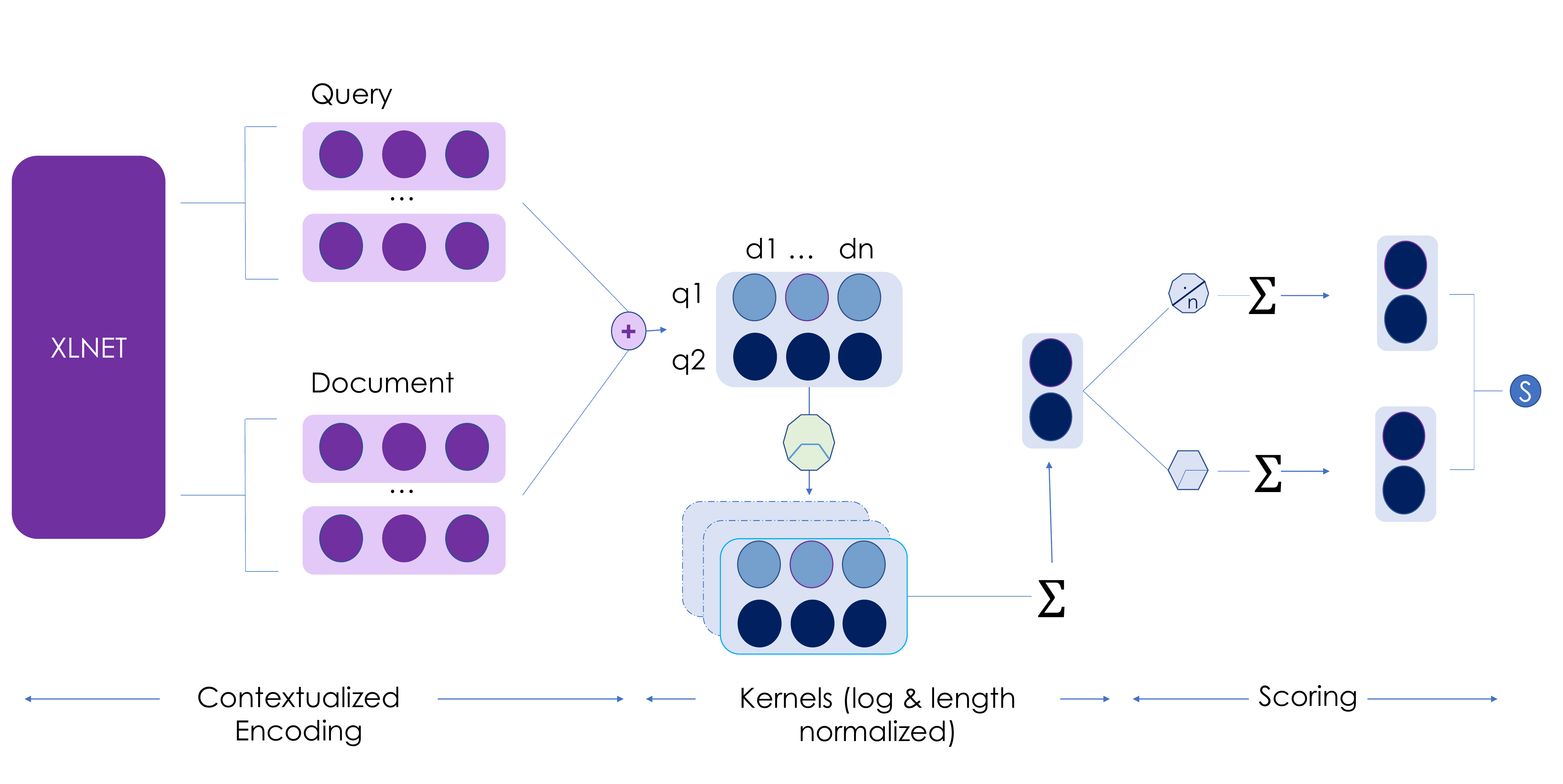}
    \caption{Creating contextualised Query-Document pairs and using them for scoring podcasts with regression.}
\end{figure*}
We use the pre-trained XLNet model (base-cased) from the HuggingFace repository. We add a linear layer on top (with dropout probability of 0.1). For training our model, We fine-tune it on a subset of the MS-Marco Passage Ranking Dataset \cite{msmarco} after freezing 5 encoder layers out of the 12 encoder layers. The subset has 100,000 samples. This linear layer is followed by a sigmoid function. This way, we are able to get a score between zero and one. The MS-Marco dataset has query-document pairs (the queries are descriptive, which works in our favour). Each query-document is labelled either 0 (irrelevant) or 1 (relevant).

We pass the query, descriptive query and the two-minute segment through the XLNet model, with separator tokens [SEP] between each of them. XLNet Regression returns a score between 0 and 1. To evaluate our model during training, we calculate metrics such as Accuracy, Precision, Recall and F1 score.

During the inference phase, we obtain the top 1000 episodes from the BM25+RM3 algorithm. From these shortlisted episodes, we create two-minute segments for each episode and index them for future use (if not indexed already).
We then re-rank the top 1000 two-minute segments using our trained model and return them as our output.

We also try out a variant of XLNet Regression, namely, XLNet Regression+Concat. It is a well-known fact that the last layer's hidden state given by the transformer encoder might not be the best representation of the text. In order to improve our results, we take the last two hidden states and pass the concatenated vector to the linear layer.

\begin{table*}[]
\centering
\begin{tabular}{|l|l|l|l|l|l|}
\hline
Model                     & P@10   & P@20   & nDCG@20 & nDCG@100 & nDCG   \\ \hline
XLNet Regression          & 0.4708 & 0.4438 & 0.3827  & \textbf{0.4501}   & \textbf{0.5414} \\ \hline
XLNet Regression + Concat & \textbf{0.4771} & \textbf{0.4542} & \textbf{0.3838}  & 0.4477   & 0.5386 \\ \hline
XLNet With Similarity     & 0.4083 & 0.3667 & 0.3083  & 0.3828   & 0.5006 \\ \hline
\end{tabular}
\caption{Results for the three models that were tested.}
\label{tab:1}
\end{table*}

\mysubsection{XLNet with Similarity}{XLNet with Similarity}
In this approach, we compute the contextualised embeddings of the query and the document as follows:
\begin{equation}
\begin{aligned}
query\;embeddings &=XLNet(query) \\
document\;embeddings &=XLNet(document)
\end{aligned}
\end{equation}
After computing these embeddings for both query and document, we find the cosine similarity matrix between the two.
\begin{equation}
    M_{ij} = cos(q_i,d_j)
\end{equation}
where $q_i$ is the $i^{th}$ query token and $d_j$ is the $j^{th}$ document token.
The shape of the matrix is $(length(query),length(document))$. The entry $M_{ij}$ corresponds to the similarity score between the $i^{th}$ query token and the $j^{th}$ document token.

After this, we obtain $k$ matrices using the $k$ RBF kernels. We choose $k$ values of $\mu$, or the mean term, and one value of $\sigma$, i.e., the standard deviation term. The purpose of doing is that each kernel focuses on a particular similarity range centred around $\mu$. $\sigma$ decides the range. This is how the computation proceeds:
 \begin{equation}
     K_{i, j}^{k}=\exp \left(-\frac{\left(M_{i j}-\mu_{k}\right)^{2}}{2 \sigma^{2}}\right)
 \end{equation}
 We sum up the matrix above along the document dimension and get a representation for each query and each kernel.
\begin{equation}
      K_{i}^{k}=\sum_{j} K_{i, j}^{k}
\end{equation}
Here, we fork the process into two paths: Log Normalisation and Length Normalisation. In Log Normalisation, we apply logarithm with base b to each query term computed above and sum them up.
\begin{equation}
    s_{\log }^{k}=\sum_{i} \log _{b}\left(K_{i}^{k}\right)    \end{equation}
Length Normalisation is performed as follows:
\begin{equation}
    s_{\mathrm{len}}^{k}=\sum_{i} \frac{K_{i}^{k}}{d_{\mathrm{len}}}
\end{equation}
Each kernel has a scalar to represent it. We use a linear layer (without the bias term) to get one common scalar for each of the log terms and the length normalisation terms.
\begin{equation}
    s_{\log }=s_{\log }^{k} W_{1} \quad s_{\text {len }}=s_{\text {len }}^{k} W_{2}
\end{equation}
The final score is computed as a weighted sum of the above two terms:
\begin{equation}
    s=s_{\mathrm{log}} * \alpha+s_{\mathrm{len}} * \beta
\end{equation}
Finally, we restricted the range to $[0,1]$ using sigmoid activation function.
\begin{equation}
    score = sigmoid(s)
\end{equation}
Again, we train the model on the MS-Marco dataset. During training, we freeze a few XLNet encoder layers. We perform inference exactly the same way as we did for XLNet Regression.

\mysubsection{Final Score Computation}{Final Score Computation}
After obtaining the BM25+RM3 score, and the score returned by either XLNet Regression or XLNet with Similarity, we compute the final relevance score as follows:
\begin{equation}
\text { Final\;Score }=\frac{ \text { X }+2\text { $\alpha$ } (\operatorname{\sigma}(B M 25+RM3)-0.5)}{1+\text { $\alpha$ }}
\end{equation}
where $\sigma$ represents the sigmoid function, $\alpha$ is a tunable scalar and $X$ is the computed XLNet Score.

This formula takes the best of both worlds: the short query, which BM25 \& RM3 are able to model well, along with the more confusing and verbose query which gets taken care of by XLNet. The purpose of applying sigmoid is that the BM25+RM3 score is not restricted to a range, which will not have both XLNet Score and BM25+RM3 score on an even keel. Since BM25+RM3 scores are positive, $\sigma(BM25+RM3)$ will lie in the range $[0.5,1]$. To expand it to the range $[0,1]$, we subtract it by $0.5$ and multiply it by $2$. 

\mysection{Experimental Study}{Experimental Study}

\mysubsection{Hyperparameters}{Hyperparameters}
We use Adam Optimiser for training our model, with a weight decay of 0.01. We experiment with our set of hyperparameters, i.e., the learning rate, the weight decay and the number of layers frozen. To find the optimal set of hyperparameters, we perform a Grid Search on the hyperparameter space. We then zero down on a learning rate of $1e^{-5}$. We try out two losses-Cross Entropy (CE) and Hinge Loss (HL). Cross Entropy Loss gives better results than Hinge Loss. We attempt to leverage information from not just the last layer of XLNet, but the previous two layers, by concatenating their embeddings. 

Due to computational constraints, we went ahead with XLNet Base. We evaluate the results using two approaches: 1) we freeze all the layers of XLNet (Frozen XLNet), 2) We keep seven layers of XLNet unfrozen (Unfrozen XLNet). Unfrozen XLNet Base gives us better results. We train our models for 5 epochs with early stopping with a patience of 2. 

\mysubsection{Results}{Results}
The results of our methods are summarized in Table \ref{tab:1}. Note that even though the performance of XLNet with Similarity is slightly lesser than XLNet Regression, it more than makes up for it with an increase in the speed of computation. Additionally, the XLNet Regression+Concat brings about an increase in the precision values and nDCG@20.

\mysection{Conclusion}{Conclusion}
In this paper, we demonstrate the ability of XLNet to perform the document retrieval task using two different approaches. XLNet outperforms BERT on most tasks, which is one of the reasons we went ahead with XLNet. In the future, we would like to evaluate the large variant of XLNet (which has 24 layers instead of 12) and a Longformers-based approach. 
We look forward to the next iteration of this competition, which might have variable-length segments to be retrieved instead of just 2-minute segments.
\bibliographystyle{ieee}
\bibliography{citations}

\end{document}